\documentstyle[twoside,fleqn,espcrc2,epsfig]{article}

\newcommand{\AmS}{{\protect\the\textfont2
  A\kern-.1667em\lower.5ex\hbox{M}\kern-.125emS}}
\newcommand{\cgs}{$erg~cm^{-2}s^{-1}$}

\title{THE BeppoSAX DEEP SURVEYS}

\author{P. Giommi$^{\rm a}$, F. Fiore
        \address{BeppoSAX Science Data Center,Rome, Italy}$^,$
        \address{Osservatorio Astronomico di Roma, Monteporzio, Italy},
        D. Ricci$^{\rm a}$,
        S. Molendi\address{IFCTR, CNR, Milano,Italy},
        M.C. Maccarone\address{IFCAI, CNR, Palermo, Italy}
        and
        A. Comastri\address{Osservatorio Astronomico di Bologna, Italy}
}
       
\begin{document}

\begin{abstract}

We present the preliminary results of a survey that makes use of several 
deep exposures obtained with the X-Ray telescopes of the BeppoSAX satellite. 
The survey limiting sensitivity is $5\times10^{-14}$ \cgs in the 2-10 keV 
band and $7\times10^{-14}$ \cgs in the harder 5-10 keV band. We find that 
the 2-10 keV LogN-LogS is consistent with that determined in ASCA surveys. 
The counts in the 5-10 keV band imply either a very hard average spectral 
slope or the existence of a population of heavily absorbed sources that can 
hardly be detected in soft X-ray surveys. 
A sample of 83 serendipitous sources has been compiled from a systematic 
search in 50 MECS images.
The analysis of the hardness ratio of this sample also implies very hard 
or heavily cutoff spectral shapes.
 
\end{abstract}

\maketitle

\section{Introduction}

Surveys of the high galactic latitude sky have greatly contributed to the 
solution of one of the most debated problems of X-ray astronomy: the nature of 
the X-ray background. It is now widely accepted that at least a substantial 
part of the apparently diffuse cosmic background is in fact due to the 
superposition of discrete extragalactic sources. 
A number of problems however still remain,
the most notable one being the so called "spectral paradox" which reflects 
the fact that the measured spectral slope of the sources that are expected to
make up the background is significantly steeper than that of the background 
itself. 
Solutions have been proposed in the framework of unified models for AGN 
which predict a large number of intrinsically obscured objects 
(Madau, Ghisellini and Fabian 1994, Comastri et al. 1995). 
These sources, however, can only be detected in sensitive hard X-ray surveys. 
Some work in this field has recently been done using ASCA data (Inoue et 
al. 1996, Georgantopoulos et al. 1997, Cagnoni, Della Ceca and Maccacaro 1998). 

Here we present 
the preliminary results of a survey that makes use of BeppoSAX imaging data.
To date, twelve high galactic latitude deep exposures have been 
carried out with the imaging instruments of the BeppoSAX satellite 
(Boella et al. 1997a) for a total exposure of over 1.1 million seconds. 
We describe the results of a detailed study of an initial sample of six  
BeppoSAX deep X-ray images. This analysis allows us to estimate a new deep 
point in the 2-10 keV LogN-LogS and to derive for the first time the source 
counts in the harder 5-10 keV band.
The results of a systematic analysis of 50 X-ray images, leading to a sizable 
sample of newly discovered hard X-ray selected sources are also presented.

\begin{figure}[htb]
\epsfig{figure=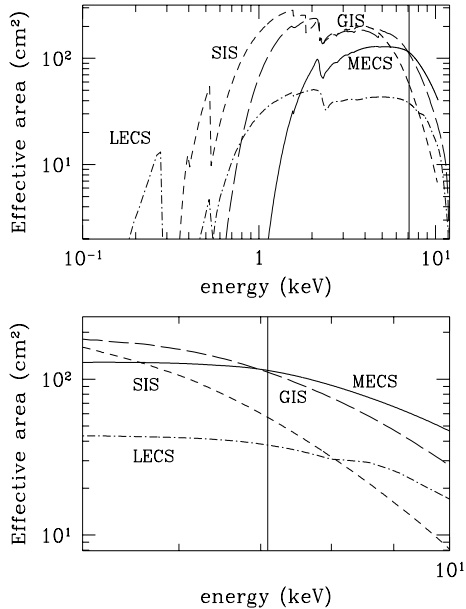, height=8.5cm, width=8.0cm}
\caption{The effective areas of the BeppoSAX and ASCA imaging 
instruments. The bottom panel shows in more detail the region 
between 5 and 10 keV}
\label{fig:one}
\end{figure}

\begin{figure}[h]
\vspace{9pt}
\epsfig{figure=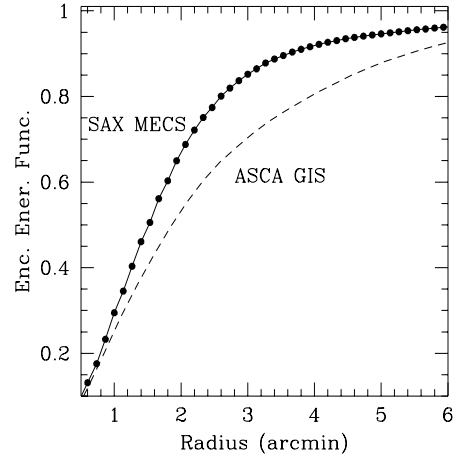, height=8.5cm, width=8.0cm}
\caption{A comparison between the BeppoSAX MECS and the ASCA GIS Point Spread 
Functions}
\label{fig:two}
\end{figure}

\begin{figure}[h]
\vspace{9pt}
\epsfig{figure=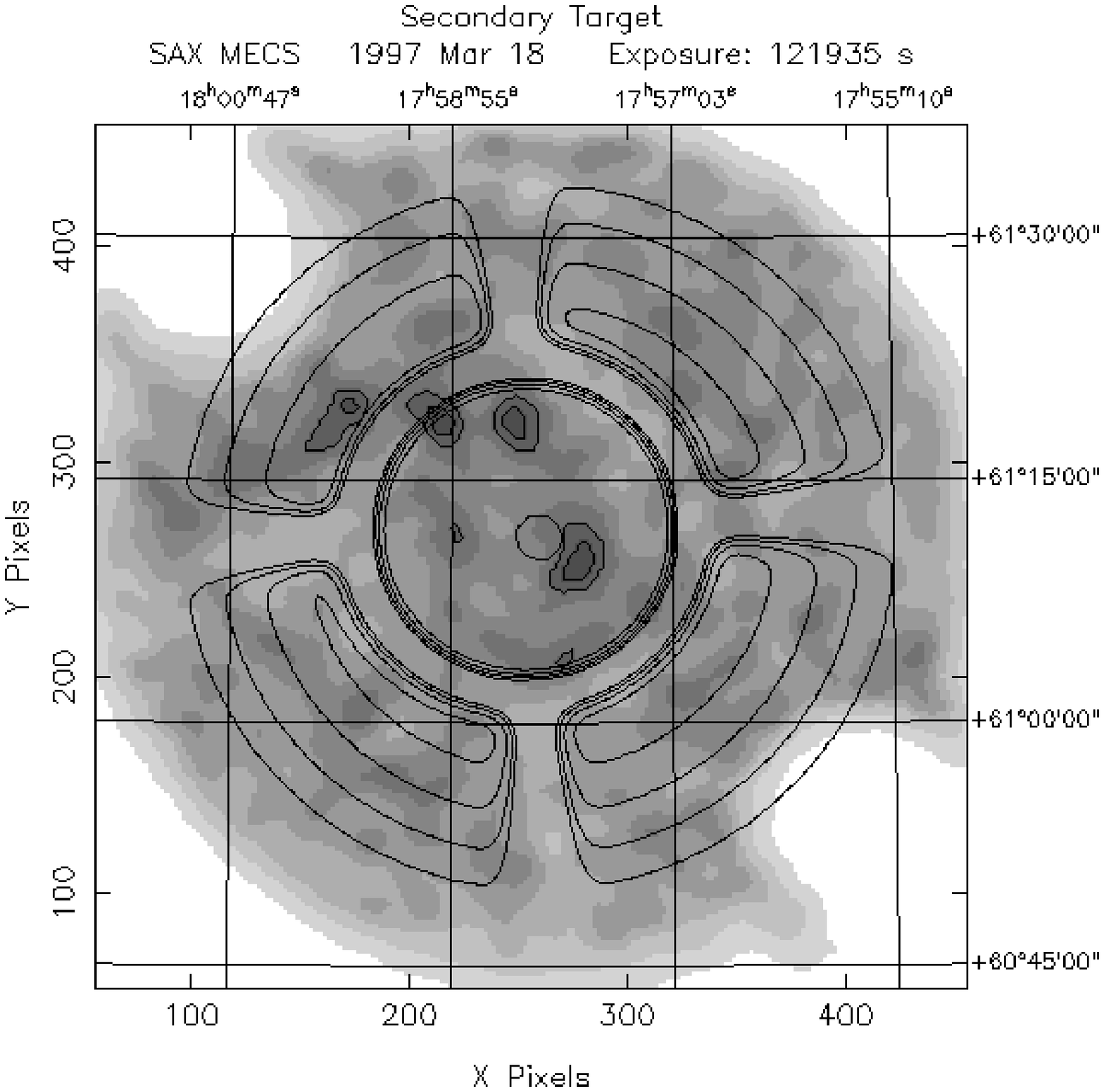, height=8.5cm, width=8.0cm, angle=-90}
\caption{One of the BeppoSAX MECS deep fields. The effects of the
window support structure are illustrated by the contour levels representing
different levels of image sensitivity}
\label{fig:three}
\end{figure}

\section{BeppoSAX X-Ray imaging capabilities}
 
   The BeppoSAX satellite carries on board four X-ray telescopes, each one 
with a Gas Scintillation Proportional Counter (GSPC) as the focal plane 
detector. 
Three of these systems, the Medium Energy Concentrator Spectrometers (MECS, 
Boella et al. 1997b) are essentially identical and operate in the 1.3-10.5 
keV band. The fourth system is known as the Low Energy Concentrator 
Spectrometer (LECS, Parmar et al. 1997) and operates between 0.1 and 10. keV 
extending the BeppoSAX spectral coverage down to the soft X-ray band.

The effective area of the LECS and MECS instruments are plotted in figure 1 
together with those of the SIS and GIS detectors of the ASCA satellite 
(Tanaka et al. 1994).

The effective area of the ASCA GIS (two units) and SIS (two units) 
is higher than that of the MECS (three units) only at low energies. 
Above 5.5 and 7 keV, the MECS instruments have a larger collecting area and 
are therefore the best instruments available at the time of writing 
for surveys in the hard band.

  One of the main advantages of the BeppoSAX imaging detectors over ASCA is the 
Point Spread Function (PSF) which is significantly sharper than that of the 
GIS or SIS, especially at high energy. 
A direct comparison is shown in figure 2 where the encircled energy function 
(for a typical spectral shape in the 2-10 keV band) of the BeppoSAX MECS is 
plotted together with that of the ASCA GIS. The radius where 50 \% of 
the photons are collected are 1.4 and 1.9 arc minutes respectively. 
The comparison rapidly worsens for ASCA at larger radii where for instance 
to collect 80 \% of the photons a radius of 3.6 arc minutes is needed 
compared to 2.6 arc minutes for BeppoSAX, corresponding to about a factor 2 in area. 
In addition, the MECS PSF in the 
soft band is largely instrument dominated and significantly improves at 
higher energies (up to about 6 keV) until scatter in the mirrors (which is 
much smaller than the spread due to the focal plane instruments at low energy) 
starts being the main contributor to the PSF. 
The difference in the response to a point source in the hard band is therefore 
even larger than that shown in figure 2. 

A significantly better PSF has the obvious advantage of improving 
sensitivity (less background has to be subtracted to the signal) and 
reduces the effects of source confusion.

In the work reported we have only used data from the MECS instruments.

\begin{figure*}[ht]
\vspace{9pt}
\epsfig{figure=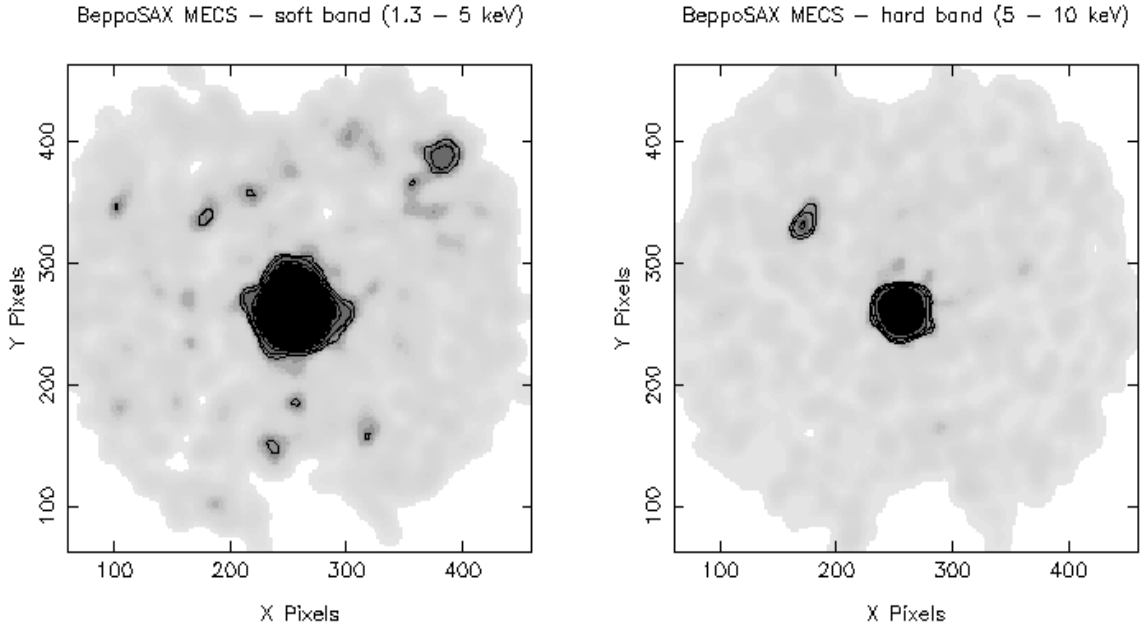, height=16.5cm, width=9.0cm, angle=-90}
\caption{Soft (1.3-5 keV) and hard (5-10 keV) MECS images of the same field. 
Note that the soft source near the edge of the FOV 
on the top right part of the plot is barely visible in the hard band, while 
the source north east of the target in the hard image is very faint in the 
1.3-5 keV band. Note also that the PSF in the hard band is significantly 
sharper than that in the 1.3-5 keV band}
\label{fig:four}
\end{figure*}

\section{Observations}

Table 1 gives the list of pointings that have been used for the analysis 
described in this paper. 
Column 1 gives the field name, column 2 the observation code in the 
BeppoSAX SDC archive, columns 3 and 4 give the Right Ascension 
and Declination (J2000.0) of the center of the field of view and column 5 
gives the exposure time in seconds.

All observations have been carried out during the Science Verification Phase 
(SVP) or as secondary observations, that is when the MECS instruments were 
collecting data 90 degrees from a Wide Field Camera (WFC, Jager et al. 1997) 
primary pointing of the Galactic center region. In some 
cases deep exposures were accumulated while the satellite was recovering from 
minor technical problems and was therefore kept in the default pointing 
direction near Polaris. 

All data used in this paper have been reprocessed using the most recently 
available reduction software. This ensures that systematic errors are kept to
a minimum, that sensitivity is slightly improved over the one that could be 
achieved with earlier data, and that source positions are estimated using 
the best attitude calibration available to date.

\begin{table*}[hbt]
\setlength{\tabcolsep}{1.5pc}
\newlength{\digitwidth} \settowidth{\digitwidth}{\rm 0}
\catcode`?=\active \def?{\kern\digitwidth}
\caption{BeppoSAX Deep Fields}
\label{tab:targetlist}
\begin{tabular*}{\textwidth}{@{}l@{\extracolsep{\fill}}cccc}
\hline
                 {Field} 
                 & {Observation\_code} 
                 & \multicolumn{1}{c}{RA} 
                 & \multicolumn{1}{c}{DEC} 
                 & \multicolumn{1}{c}{Exposure time} \\
\hline
WFC Secondary & 00001003 & 17 30 42. & +60 55 33 & 65841\\
WFC Secondary & 00001005 & 16 33 43. & +59 44 13 & 97937\\
WFC Secondary & 00001007 & 18 04 03. & +61 08 43 & 116720 \\
WFC Secondary & 00001008 & 17 58 15. & +61 10 19 & 118969\\
WFC Secondary & 00001009 & 18 19 50. & +60 56 37 & 114574\\
Polaris 1     & 00001011 & 01 38 18. & +89 13 51 & 115498 \\
\hline
\end{tabular*}
\end{table*}

\section{Data Analysis}

To search for faint sources in deep BeppoSAX X-ray images we have adopted  
a source detection procedure that is a variation of the detect routine 
included in the XIMAGE package (Giommi et al. 1991). 
The method consists in first convolving the X-ray image with a function 
of size similar to that of the MECS PSF (a wavelet function 
in this case) to smooth the image and increase contrast. 
As a second step a standard slide-cell detection method is used to locate 
count excesses above the local background.
The statistics of each candidate detection is then accurately studied and the 
final net counts are estimated from the original (un-smoothed) image 
to preserve Poisson statistics. 
The background is calculated in a number of source-free boxes 
near the position of the candidate detection and is rescaled taking into 
account of spatial variations of the MECS background. To maximize sensitivity 
the size of the box used for flux estimation is chosen to be that where 
the signal to noise ratio is expected to be maximum given the local background,
source intensity and local PSF.

Extensive Monte Carlo simulations made using the BeppoSAX on-line simulator, 
show that this source detection method is fast and highly sensitive to point 
sources even in presence of some source confusion. 
The method is also statistically robust since final acceptance 
and flux estimation on each detection is simply based on Poisson statistics  
and does not require any time consuming fitting procedures which, especially 
in case of weak signal, may not always be stable.

\begin{figure}[htb]
\vspace{9pt}
\epsfig{figure=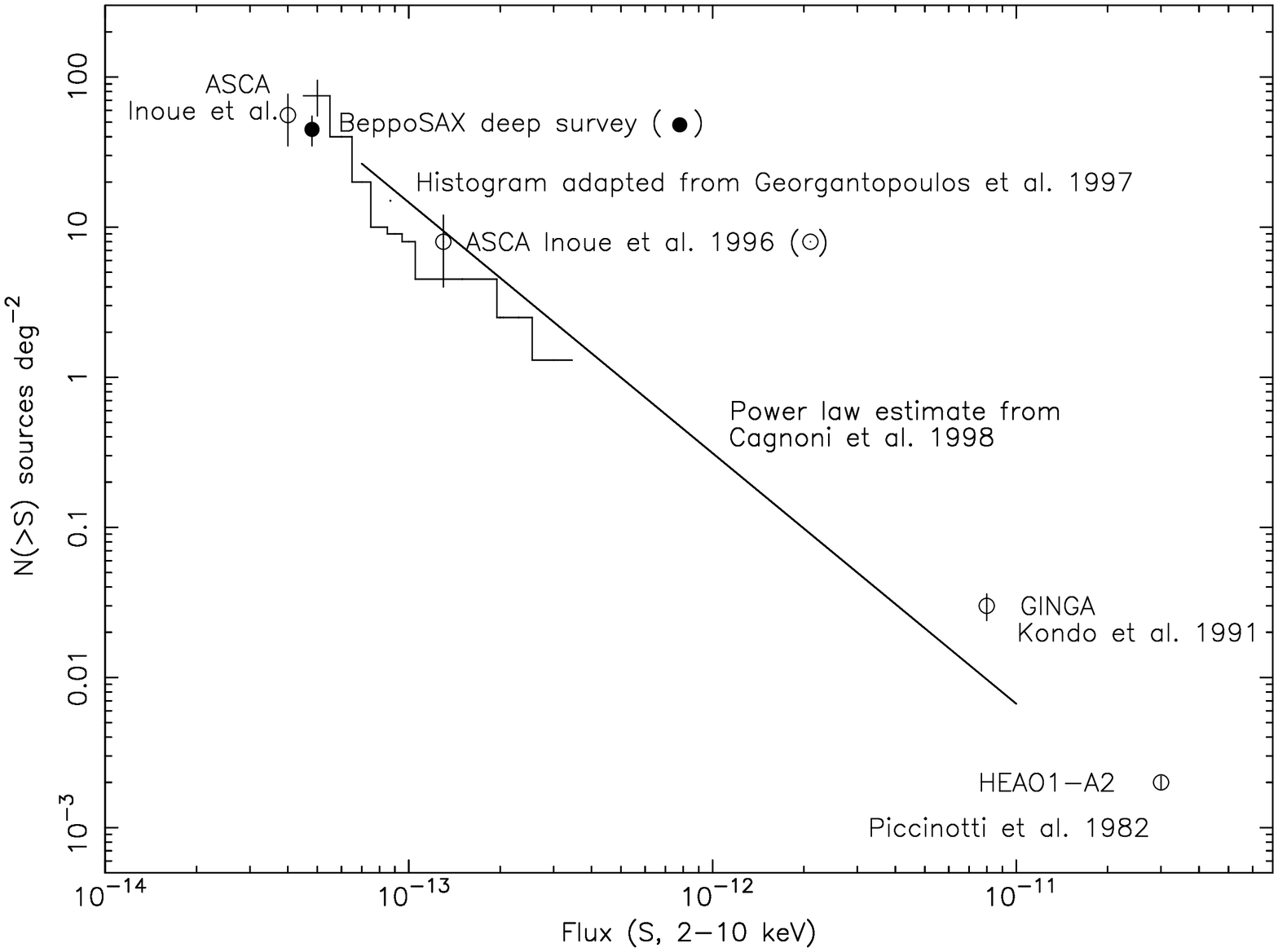, height=8.5cm, width=8.0cm, angle=-90}
\caption{The 2-10 keV LogN-LogS as estimated with various X-ray astronomy 
satellites. The BeppoSAX deep survey point at S=$5\times10^{-14}$ \cgs is 
consistent with other measurements.}
\label{fig:five}
\end{figure}
 
\subsection{Source detection in the 2-10 keV band}

  The MECS X-Ray images have a circular field of view with radius of 
approximately 25 arcminutes. 

The sensitivity to X-ray sources is a fairly strong function of 
the position in the image. Several factors contribute to this complex 
dependency. 
Figure 3 shows one of the deep fields used in our survey shown together 
with its sensitivity map overlaid as a set of contour levels. 
The most notable feature is the obscuration due to the detector window support 
structure (also known as the strongback) that is clearly visible in the 
sensitivity contours.
The strongback is made of 600 microns of Beryllium and absorbs most of the 
X-rays below 5 keV.
The two small empty regions near edge of the field of view on the top left 
and bottom right part of the image, are due to the removal of the area 
affected by the on-board calibration sources. 
Another cause of the sensitivity dependency on distance from the center 
is the well known vignetting effect which for the MECS detectors can 
be as high as a factor 3.5 at very large off axis angles. 
Finally the PSF also degrades with distance from the center; however, since 
the spread in the MECS is dominated by detector effects (at least up to 
about 6 keV), this is a smaller problem than in most other X-ray telescopes.

To avoid complications due to complex variations in sensitivity across the 
field of view for our survey we only consider the central region of 9 
arcminutes radius where sensitivity does not vary more than a few 
percent (see figure 3). 
Full exposure map corrections were applied to all images considered.
The exposure in all fields listed in table 1 is sufficiently long to 
guarantee a theoretical sensitivity (in absence of source confusion) of
$2-3\times10^{-14}$ \cgs. However, because of source confusion 
problems (see below) we decided to limit our survey to a sensitivity of 
$5\times10^{-14}$ \cgs. 
Each field therefore contributes to our survey with a constant
area equal to 0.07 square degrees with fixed sensitivity of 
$5\times10^{-14}$ \cgs.

Images were accumulated in the 2-10 keV band for all the six pointings 
listed in table 1 and the detection algorithm described above was run 
on all data. 

Sources were accepted according to the following criteria.

1) The off-axis angle is less or equal to 9 arcminutes;

2) Statistical probability that the source is a fluctuation of the local 
background is less than $1\times 10^{-4}$. This implies that less than one 
spurious source is expected in the entire survey.

3) The count rate corresponds to a 2-10 keV flux higher or equal to 
$5\times10^{-14}$ \cgs assuming a power law spectrum with an energy slope 
of 0.7.

The analysis of all BeppoSAX deep fields was also carried out on the
soft (1.3-5 keV), hard (5-10 keV) band. This ensures that sources with 
very soft or hard spectrum (figure 4) are 
not missed. 

\section{The 2-10 keV LogN-LogS}

Since the sensitivity limit was fixed to the same value in all images 
the estimation of the source counts in our survey is simply a single point
in the LogN-LogS plane.

Nineteen sources were detected in the 0.42 square degrees of the survey
corresponding to a density of $45\pm 10$ sources/sq degree with flux
equal or larger than $5\times10^{-14}$ \cgs. 

At present most of these sources are unidentified. 
Figure 5 shows the MECS deep fields point in the 2-10 keV LogN-LogS plane 
together with the results of  ASCA (Georgantopoulos et al. 1997, 
Cagnoni et al. 1998), Ginga (Kondo et al. 1991) and HEAO1 (Piccinotti 
et al. 1982). The BeppoSAX counts are clearly in good agreement with 
the results of ASCA.

\section{Simulations and the source confusion problem}

To address the problem of source confusion and biases at faint fluxes near 
the detection limit (see e.g. Hasinger et al. 1998), we simulated a large 
number of MECS deep fields using the BeppoSAX on-line data simulator described 
below. 

\subsection{The BeppoSAX simulator}

The BeppoSAX on-line X-ray data simulator (http://www.sdc.asi.it/simulator) 
is a facility provided by the BeppoSAX Science Data Center (Giommi \& 
Fiore 1997) that can be used to make detailed simulations of LECS and MECS 
X-ray imaging data.
This software generates X-ray sources randomly distributed across the image 
following a LogN-LogS equal to that determined using ASCA data by Cagnoni 
et al (1998). Detector effects including position and energy dependent PSF 
and telescope vignetting, are fully taken into account. The format of the 
output files is identical to that of LECS or MECS science archive data so 
that the same analysis software and procedures can be used for simulated 
and real data.

\subsection{Simulation runs}

Fifty deep fields with exposures of 100,000 seconds each, were generated 
and subsequently analyzed following the same method used 
for the survey. This resulted in the detection of a relatively large sample 
of sources which was then used to estimate the source counts at different 
flux limits. These estimates were finally compared to the LogN-LogS used as 
input for the simulation. 

No significant bias in the estimation of the LogN-LogS could be found 
down to a flux of approximately $5\times10^{-14}$ \cgs.
At lower flux levels several sources could still be detected but in this regime 
source confusion causes severe problems in the determination of source flux
and accurate positions (see also the results of Hasinger et al. 1998).

\section{Hard sources}

The results in the 2-10 keV band encouraged us to extend our 
analysis of BeppoSAX deep fields to the hard band and to estimate
the source counts between 5 and 10 keV. 
The BeppoSAX MECS are well suited for this determination since 
their effective area peaks at around 5-6 keV (see figure 1) and
the PSF in the hard band is sharper than that in the full 2-10 keV band. 
Source confusion is not a major problem anymore.

\subsection{The hard (5-10 keV) LogN-LogS}

We have repeated the analysis carried out for the 2-10 keV data 
on images accumulated in the harder 5-10 keV band.
The sensitivity limit, set equal for all six images as in the case of the 
2-10 keV band, was $7\times10^{-14}$ \cgs. Because of possible small 
systematic effects and of the uncertainty in the knowledge of the 
sources spectral slope we conservatively associate an uncertainty of 
$1\times10^{-14}$ in the flux limit.

This analysis has led to the detection of 8 sources yielding a density 
of $19\pm 7$ sources/sq degree.
  
Figure 6 plots the BeppoSAX (5-10 keV) source density at 
$7\times10^{-14}$ \cgs in the LogN-LogS plane. 
Since no results from other satellites are available yet in this band, 
comparison with other data can only be done by extrapolating  
results obtained in softer bands assuming different spectral slopes.  
The solid line shown in figure 6 represents the 5-10 keV LogN-LogS as 
expected from the Cagnoni et al (1998) counts assuming a spectral 
slope of $\alpha = 0.4 $. The BeppoSAX counts are about a factor 2
higher than expected, indicating that a very flat ($\alpha \le 0.2$), 
and most probably unplausible, average source spectrum must be assumed. 
Alternatively, a population of heavily cutoff sources could explain the 
difference.

The still relatively large statistical uncertainties, however do not allow
strong conclusions and more data are necessary to confirm these early results.
The analysis of all existing BeppoSAX deep fields will contribute to 
address this point.

The nature of the BeppoSAX hard X--ray sources is still unknown.
An important contribution is expected from obscured type 2 
objects as predicted in the AGNs synthesis models for the 
hard X--ray background (Madau Ghisellini and Fabian 1994, 
Comastri et al. 1995). These sources can hardly be detected
in the soft X--rays even at the faintest ROSAT limit
while in harder energy bands they should become detectable.
To further investigate this possibility we have computed
the source space density in the 5--10 keV band predicted by the
Comastri et al. (1995) model (figure 7).
The dotted line represent the contribution of unabsorbed
type 1 objects (e.g. Seyfert 1 galaxies and quasars), while the
solid line gives the summed contribution of unabsorbed and absorbed 
objects. The model predictions are consistent (at the 2$\sigma$ level) 
with the BeppoSAX observations.

The small discrepancy between the observed and predicted source surface
density may be due to several factors, including poor statistics.
The model predictions have been computed assuming for the AGN luminosity
function and cosmic evolution the best fit parameters derived by ROSAT
(Boyle et al. 1993), which may not be appropriate especially for the 
highly absorbed sources.
The contamination of non--AGN sources (which are not included
in the model) to the observed counts could be important given the 
relatively low number of hard X--ray sources (eight).
A different explanation would imply the contribution of a new class
of very hard X--ray sources which have been missed by previous surveys.
A more detailed discussion of these points is deferred to a future
paper.

\section{Hardness ratio analysis}

Figure 4 shows the soft (1.3-4.0 keV) and hard (4-10 keV) X-ray image of 
a MECS observation where two serendipitous sources with different spectral
shape have been detected. The source near the edge at the top right 
of the soft image is only detected in the soft band, while most of the 
photons from the source to the northeast of the target are detected above 
4 keV.

Similarly, the sources that we have detected in our survey show a variety 
of spectral slopes, ranging from soft stellar like spectra to very hard 
ones where nearly all the photons are detected in the hard band.

To study this spectral variety in a quantitative way we have conducted a 
systematic search for serendipitous sources in a sample of 50 
MECS images (see Fiore et al. 1998) selected among SVP observations 
or from fields where 
the SAX team or one of the authors of this work is the PI. 
This search led to the detection of 83 serendipitous sources. 
For each detection we have estimated the hardness ratio (defined as 
HR=(S-H)/(S+H), where S are the net counts in the 1.3-4.5 keV band and 
H are the counts in the 4.5-10. keV band).

Figure 8 plots the hardness ratios as a function of source total count-rate. 
The solid triangles represent the serendipitous sources discussed above, 
(the error bars are not plotted but typically are of the order of 0.2-0.3). 
Open circles represent a set of pointed AGN (observed during SVP or that 
are AO1 observations where one of the authors of this work is a PI or one 
of the main co-investigators); filled circles are pointed sources which 
are well known to be intrinsically absorbed sources.  
The scale on the right vertical axis shows the equivalent (energy) spectral 
slope for an assumed power law spectrum. We see that the hardness ratio 
of the serendipitous sources is systematically lower than that of pointed 
(and unabsorbed) AGN, with typical values corresponding to very flat spectral 
slopes ($\alpha $ around 0!).   
The fact that there are very few serendipitous sources
with hardness ratio similar to that of the pointed AGN is a simple selection 
effect: at very low count rates, near the detection limit, 
a soft source would not be detected in the hard band and therefore no 
hardness ratio for it can be estimated. 
From figure 8 we can conclude that a substantial population of very hard
(or very absorbed) serendipitous sources has been detected, consistently with 
the expectations from the LogN-LogS results.
\begin{figure}[htb]
\epsfig{figure=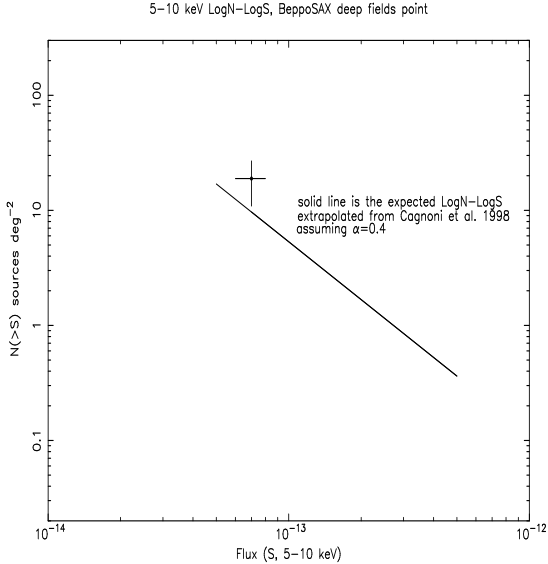, height=8.cm, width=8.0cm, angle=-90}
\caption{The 5-10 keV BeppoSAX counts compared to the expectations from
the ASCA LogN-LogS assuming an average (energy) spectral slope 
$\alpha = 0.4 $}
\label{fig:toosmall}
\end{figure}

\begin{figure}[htb]
\vspace{9pt}
\epsfig{figure=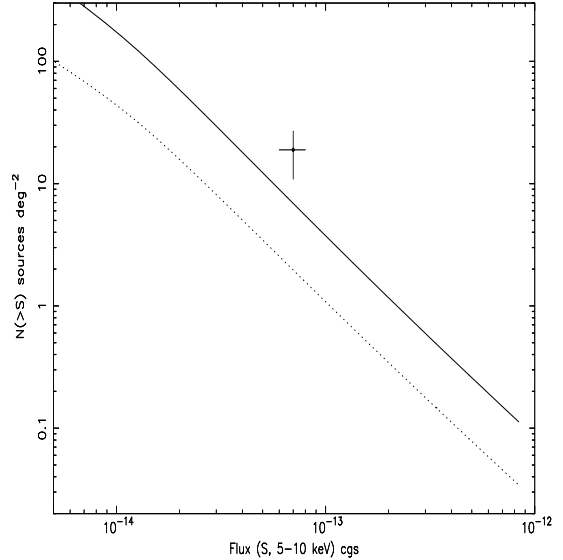, height=8.cm, width=8.5cm,angle=-90}
\caption{The prediction of the Comastri et al (1995) model (solid line) 
compared to the 5-10 keV source counts from the BeppoSAX deep survey.
The dotted line represents the contribution of unabsorbed AGN.}
\label{fig:fig8}
\end{figure}

\begin{figure}[htb]
\vspace{9pt}
\epsfig{figure=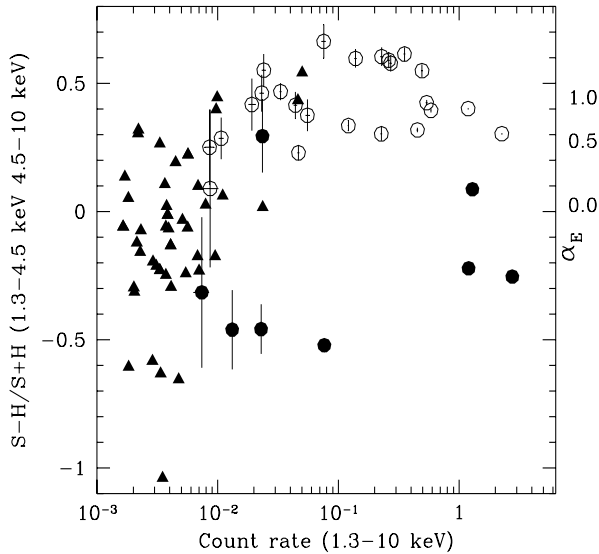, height=8.5cm, width=8.0cm}
\caption{The hardness ratios of MECS serendipitous sources (filled triangles)
plotted as a function of count rate. For comparison the HR of several pointed
AGN (open circles) and highly absorbed targets (filled circles).}
\label{fig:largenenough}
\end{figure}

\section{Conclusions}

We have presented early results from a project that deals with a
relatively deep survey of the high galactic latitude 
sky in the 2-10 keV and 5-10 keV bands. The data used are six deep 
observations made with the imaging instruments of the BeppoSAX satellite.

The results in the 2-10 keV band are in good agreement with those found in the 
ASCA deep surveys. This confirms that at a flux of $5\times10^{-14}$ \cgs 
about 30\% of the cosmic background is 
resolved in point sources and that there is an excess of hard X-ray sources 
compared to the predictions derived from the ROSAT 0.5-2. keV LogN-LogS
(Hasinger et al. 1993, Inoue at al. 1996, Geogantopulous et al. 1997). 
This last conclusion is further strengthened by the high number of sources  
found in the hard 5-10 keV band. 
This result can be explained assuming the existence of a population of objects 
that could not be detected by ROSAT but is contributing to the BeppoSAX 
space density. 
Several hard sources have indeed been detected by us as serendipitous 
sources in a set of 50 MECS X-ray images.

The source surface density at the BeppoSAX flux limit 
is in overall agreement with the expectation of the AGN synthesis
models for the XRB. 

The population of hard sources could include Seyfert 2 galaxies with high 
intrinsic absorption 
(NH around $1\times 10^{23}$), as predicted by unified models for AGN 
(e.g. Antonucci 1993) and consistent with BeppoSAX direct observations 
(Salvati et al. 1997).
A significant test for these models would be the comparison with 
the optical identifications of the BeppoSAX hard X--ray sources. 
Deeper observations with the next generation of 
X--ray observatories with good angular resolution and high spectral 
sensitivity in the hard X--ray band (i.e. AXAF and XMM)
will also allow to tighten the constraints on the nature of the weak 
hard X--ray sources and in turn provide a definitive answer on the
origin of the X--ray background.

\section{Acknowledgments}

We thank T. Maccacaro, and M. Salvati for providing some of their 
BeppoSAX results in advance of publication.

\end{document}